\documentclass[twocolumn]{aastex62}
\usepackage{natbib}
\usepackage{epsfig}
\usepackage{graphicx}
\usepackage{subfigure}
\usepackage{float}
\usepackage{amsmath}
\usepackage{color}
\usepackage{amssymb}
\usepackage{amsfonts}
\usepackage{apjfonts}

\newcommand{\Mpc}{\mathrm{~km~s^{-1}~Mpc^{-1}}}

\def\({\left(}
\def\){\right)}
\def\[{\left[}
\def\]{\right]}

\shortauthors{Liu et al.}

\begin{document}

\title{Revisiting the Hubble constant, spatial curvature and cosmography with strongly lensed quasar and Hubble parameter observations}

\correspondingauthor{Shuo Cao}
\email{caoshuo@bnu.edu.cn}

\author{Tonghua Liu}
\affiliation{School of Physics and Optoelectronic, Yangtze University, Jingzhou, 434023, China}
\affiliation{Frontier Science Institute of Astronomy and Astrophysics, Beijing Normal University, Beijing, 100875, China}

\author{Shuo Cao$^\ast$}
\affiliation{Frontier Science Institute of Astronomy and Astrophysics, Beijing Normal University, Beijing, 100875, China}
\affiliation{Department of Astronomy, Beijing Normal University, 100875, Beijing, China}

\author{Marek Biesiada}
\affiliation{National Centre for Nuclear Research,
Pasteura 7, 02-093 Warsaw, Poland}

\author{Shuaibo Geng}
\affiliation{National Centre for Nuclear Research,
Pasteura 7, 02-093 Warsaw, Poland}

\begin{abstract}
It is well known that strong lensing time-delays offer an opportunity of a one-step measurement of the Hubble constant $H_0$ which is independent of cosmic distance ladder. In this paper, we go further and propose a cosmological model-independent approach to simultaneously determine the Hubble constant and cosmic curvature with strong lensing time-delay measurements, without any prior assumptions regarding the content of the Universe. The data we use comprises the recent compilation of six well studied strongly lensed quasars, while the cosmic chronometer data are utilized to reconstruct distances via cosmographic parameters. In the framework of third-order Taylor expansion and (2, 1) order Pad\'{e} approximation for for cosmographic analysis, our results provides
model-independent estimation of the Hubble constant $H_0 = 72.24^{+2.73}_{-2.52} \Mpc$ and $H_0 = 72.45^{+1.95}_{-2.02} \Mpc$, which is well consistent with that derived from the local distance ladder by SH0ES collaboration. The measured cosmic curvature $\Omega_k=0.062^{+0.117}_{-0.078}$ and $\Omega_k=0.069^{+0.116}_{-0.103}$ shows that zero spatial curvature is supported by the current observations of strong lensing time delays and cosmic chronometers. Imposing the prior of spatial flatness leads to more precise (at 1.6$\%$ level) determination of the Hubble constant $H_0=70.47^{+1.14}_{-1.15} \Mpc$ and $H_0=71.66^{+1.15}_{-1.57} \Mpc$, a value located between the results from \textit{Planck} and SH0ES collaboration. If a prior of local (SH0ES) $H_0$ measurement is adopted, the curvature parameter constraint can be further improved to $\Omega_k=0.123^{+0.060}_{-0.046}$ and $\Omega_k=0.101^{+0.090}_{-0.072}$, supporting no significant deviation from a flat universe. Finally, we also discuss the effectiveness of Pad\'{e} approximation in reconstructing the cosmic expansion history within the redshift range of $z\sim2.3$, considering its better performance in the Bayes Information Criterion (BIC).

\end{abstract}

\keywords{Strong gravitational lensing (1643) ---  Cosmological parameters (339)}

\section{Introduction}
In  modern cosmology,  the spatially flat model with cosmological constant plus cold dark matter ($\Lambda$CDM)  has withstood the tests against most of
the observational evidences \citep{2007ApJ...659...98R,2017PhRvD..95b3524A,2016A&A...594A..13P,2011MNRAS.416.1099C,2012JCAP...03..016C,2014PhRvD..90h3006C,2015ApJ...806..185C}. However, the standard cosmological model suffers some unresolved issues, such as, the Hubble tension \citep{2017NatAs...1E.169F,2021APh...13102605D}, the cosmic curvature problems \citep{2020NatAs...4..196D}, well-known \textit{fine-tuning} and cosmic \textit{coincidence problem} \citep{1989RvMP...61....1W,2000astro.ph..5265W,2001PhRvL..87n1302D,2010ApJ...711..439C,2011MNRAS.416.1099C,2018RAA....18...66Q}. Especially, there is $4.4\sigma$ tension  between the Hubble constant ($H_0$) inferred within $\Lambda$CDM from cosmic microwave background (CMB) anisotropy (temperature and polarization) data \citep{2020A&A...641A...6P} and its value measured through Cepheid - calibrated distance ladder by the \textit{Supernova $H_0$ for the Equation of State} collaboration (SH0ES) \citep{2019ApJ...876...85R}. This inconsistence may be  caused by unknown systematic errors in astrophysical observations or reveal new physics significantly different from $\Lambda$CDM. Recently, some works suggested that the $H_0$ tension could possibly be caused by the inconsistency of spatial 
curvature between the early-universe and late-universe \citep{2021APh...13102605D,2020NatAs...4..196D,2021PhRvD.103d1301H}. More specifically, combining the \textit{Planck} temperature and polarization power spectra data, the work showed that a closed Universe $\Omega_k=-0.044^{+0.018}_{-0.015}$ was supported, as a consequence of higher lensing amplitude \citep{2020A&A...641A...6P} supported by the data. However, combination of the Planck lensing data and low redshift baryon acoustic oscillations observations, leads to the preference of a flat Universe with curvature parameter precisely constrained within $\Omega_k=0.0007\pm0.0019$. It should be emphasized that both methods 
invoke a particular cosmological model --- the non-flat $\Lambda$CDM. To better understand the inconsistence between the Hubble constant and the curvature of Universe inferred from local measurements and \textit{Planck} observations, it is necessary to seek for the
cosmological model-independent method that constrains both $H_0$ and $\Omega_k$.

Strong gravitational lensing (SGL) systems, which already demonstrated their constraining abilities on dark matter density \citep{Cao11b,Cao12a}, dark energy equation of state \citep{Cao12b,LiuT19}, and the velocity dispersion function of early-type galaxies \citep{Ma19,Geng21}, provide a valuable opportunity for model-independent constraints on both $H_0$ and $\Omega_k$. Firstly, from the measurement of time-delay between multiple images from SGL, one can measure the Hubble constant effectively and directly. For details, practical approaches and results of the Hubble constant measurements using SGL time-delay one can refer to  \citep{2020ApJ...895L..29L,2019ApJ...886L..23L,2020ApJ...900..160L,2019A&A...628L...7T}. Secondly,  combining the time-delay measurements with other SGL observations and using the well-known distance sum rule, one can simultaneously determine the spatial curvature and the Hubble constant without adopting any particular cosmological model. Recently, the work by  \cite{2019PhRvL.123w1101C} first presented such methodology to simultaneously constrain the spatial curvature and the Hubble constant with SN Ia observations and SGL time-delay measurement. In further research, \citet{2020ApJ...897..127W} extended this approach by considering the non-linear relation between ultraviolet and X-ray fluxes observations of high redshift quasars. However, the SGL time-delay measurements applied to cosmology require three angular diameter distances. Hence, observations using angular diameter distances are more intuitive and efficient in this context, and following this path \citet{2021MNRAS.503.2179Q} further determined the $H_0$ and $\Omega_k$ by combining SGL time-delays with compact structure in radio quasars acting standard rulers, and obtained stringent results $H_0=78.3\pm2.9\Mpc$ and $\Omega_k=0.49\pm0.24$. However, one should emphasize here that whether using quasars as standard rulers or supernovae as standard candles, the problem of relative distance calibration is unavoidable. It should be done without invoking any particular background cosmology model in order not to fall into a circularity.
One general approach often used by the researchers is to take a polynomial expansion of the cosmological distance in redshift (or certain function of redshift like $\log{(1+z)}$). Such reconstruction procedure often involves questions about the physical meaning of polynomial coefficients.

In this paper, we will determine the curvature parameter and the Hubble constant simultaneously using the direct measurements of Hubble expansion rates provided by the proposed differential ages of passively evolving galaxies (also known as cosmic chronometers) \citep{2002ApJ...573...37J}, focusing on six SGL time-delay measurements released by the $H_0$ Lenses in COSMOGRAIL's Wellspring (H0LiCOW) collaboration \citep{2020MNRAS.498.1420W}. For the purpose of distance reconstruction, instead of the polynomial expansion, we will use a method named \textit{cosmography} \citep{2019PhRvL.123w1101C,2020ApJ...897..127W,2021MNRAS.503.2179Q}.
Examples of using cosmographic approach can be found in e.g.
\citep{2015CQGra..32m5007V,2019IJMPD..2830016C,2016PhRvD..94h3525D,2019NatAs...3..272R,2015ApJ...815...33R}. The advantage of cosmography consists in minimal prior assumptions about the Universe (only its homogeneity and isotropy) and in using the expansion of the Hubble function around the present time, which is much more physical than polynomial expansion of distance. This paper is organized as follows: In Sec. 2, we present the methodology and observational data. The results and discussion are given in Sec. 3. Finally, we summarize our findings in Sec. 4.

\section{Methodology and observations} \label{sec:data}

For a general homogeneous and isotropic universe, the FLRW metric reads
\begin{equation}
ds^2=c^2dt^2-\frac{a(t)^2}{1-kr^2}dr^2-a(t)^2r^2d\Omega^2,
\end{equation}
where $a(t)$ is the scale factor and $c$ is the speed of light, and $k$ is dimensionless curvature taking one of three values $\{-1, 0, 1\}$. The cosmic curvature parameter $\Omega_k$ is related to $k$ and the Hubble constant $H_0$, as $\Omega_k=-c^2k/a^2_0H_0^2$. Having in mind SGL system where three elements occur: the source, the lens and the observer, let us introduce dimensionless comoving distances $d_l\equiv d(0, z_l)$, $d_s\equiv d(0, z_s)$ and $d_{ls}\equiv d(z_l, z_s)$ where the cosmological length scale $c/H_0$ is factored out. As a specific example, the dimensionless comoving distance between the lensing galaxy (at redshift $z_l$) and the background source (at redshift $z_s$) is given by
\begin{eqnarray}
\nonumber d(z_l, z_s)&=& \frac{H_0}{c}(1+z_s) D_A(z_l,z_s)\\
&=&\frac{1}{\sqrt{|\Omega_k|}}f\left(
\sqrt{|\Omega_k|}\int^{z_s}_{z_l}\frac{dz'}{E(z')} \right),
\end{eqnarray}
where
\begin{equation}\label{eq3}
f(x)=\left\{
   \begin{array}{lll}
   \sin(x)\qquad \,\ \Omega_k&<0, \\
   x\qquad\qquad \,\ \Omega_k&=0, \\
   \sinh(x)\qquad \Omega_k&>0,  \\
   \end{array}
   \right.
\end{equation}
and $D_A(z_l,z_s)$ represents the angular diameter distance from $z_l$ to $z_s$,  which is given by $D_A(z_l,z_s)=\frac{c}{1+z_s}\int_{z_l}^{z_s}\frac{dz'}{H(z')}$ and the expansion rate at redshift $z$ is: $H(z) = H_0 E(z)$.
In the framework of FLRW metric, one can use the well-known distance sum rule (DSR)  to relate these three distances in non-flat Universe \citep{2015PhRvL.115j1301R}
\begin{equation}\label{eq4}
d_{ls}={d_s}\sqrt{1+\Omega_kd_l^2}-{d_l}\sqrt{1+\Omega_kd_s^2}.
\end{equation}
Note that in terms of dimensionless comoving distances DSR will reduce to an additivity relation $d_s =d_l+d_{ls}$ in the flat universe ($\Omega_k=0$). The original idea of cosmological application of the distance sum rule in general, and with respect to gravitational lensing data in particular can be traced back to \citep{2015PhRvL.115j1301R}.
Moreover, one can rearrange Eq.~(4) as \begin{equation}\label{eq5}
\frac{d_ld_s}{d_{ls}}=\frac{1}{\sqrt{1/d_l^2+\Omega_k}-\sqrt{1/d_s^2+\Omega_k}},
\end{equation}
which provides a model-independent test for the Copernican principle \citep{Cao2019,2019PhRvD.100b3530Q,ZhangY22} and the viscosity of dark matter \citep{Cao21,Cao22}, with different measurements of cosmic curvature and
dark matter viscosity at different redshifts. The expression (\ref{eq5}) multiplied by $1+z_l$ and with dimensionality of length recovered is known as the time delay distance $D_{\mathrm{\Delta t}}$.

\textit{Time-delay distances from lensed quasars.---}
In strong lensing systems with quasars acting as background sources, the time difference (time-delay) between two images of the source, which could be  measured from variable AGN light curves, depends on the time-delay distance $D_{\mathrm{\Delta t}}$ and the geometry of the Universe as well as the gravitational potential of the lensing galaxy \citep{1990CQGra1319P,1990CQGra1849P}
\begin{equation}
\Delta t_{i,j} = \frac{D_{\mathrm{\Delta t}}}{c}\Delta \phi_{i,j}(\mathbf{\xi}_{lens}),
\label{relation}
\end{equation}
where $c$ is the speed of light, and $\Delta\phi_{i,j}=[(\mathbf{\theta}_i-\mathbf{\beta})^2/2-\psi(\mathbf{\theta}_i)-(\mathbf{\theta}_j-
\mathbf{\beta})^2/2+\psi(\mathbf{\theta}_j)]$
is the Fermat potential difference between the image $i$ and image $j$ which is determined by the lens model parameters $\mathbf{\xi}_{lens}$ inferred from high resolution imaging of the host arcs. It is worth noting that the line-of-sight (LOS) mass structure could also affect the Fermat potential inference. The contribution of the line-of-sight mass distribution to the lens requires deep and wide field imaging of the area around the lens system. $\mathbf{\theta}_{i}$ and $\mathbf{\theta}_{j}$ are angular positions of the image $i$ and $j$ in the lens plane \citep{2016A&ARv..24...11T}. The two-dimensional lensing potential at the image positions, $\psi(\mathbf{\theta}_i)$ and $\psi(\mathbf{\theta}_j)$, and the unlensed source position $\mathbf{\beta}$ can be determined by the mass model of the system \citep{2015A&A...580A..38R}. The time-delay distance $D_{\mathrm{\Delta t}}$ is a combination of three angular diameter distances expressed as
\begin{equation}
D_{\mathrm{\Delta
t}}\equiv(1+z_l)\frac{D^A_{l}
D^A_{s}}{D^A_{ls}}=\frac{c}{H_0}\frac{d_ld_s}{d_{ls}},
\end{equation}
where the superscript ($A$) denotes angular diameter distance. One can clearly see that the distance ratios $d_ld_s/d_{ls}$ can be assessed from the
observations of time delay in SGL systems.
Meanwhile, if the other two dimensionless distances $d_l$ and
$d_s$ can be obtained from observations, the measurement of
$\Omega_k$ and $H_0$ could be directly obtained with time-delay
measurements of $\Delta t_{i,j}$ and well-reconstructed lens potential difference of $\Delta \phi_{i,j}$ between two images. See \citet{
2019ApJ...871..113L,2020MNRAS.498.2871H,2021MNRAS.503.1096D,2021A&A...656A.153S,2021ApJ...906...26L,2022ApJ...927..191B} for more recent works on the extensive applications of SGL time delays.

In the milestone paper of \citet{2020MNRAS.498.1420W}, the six lensed quasars sample was jointly used to measure the $H_0$ by H0LiCOW collaboration under six dark energy cosmological models.  The six lenses with measured time delays are B1608+656 \citep{2010ApJ...711..201S,2019Sci...365.1134J}, RXJ1131-1231 \citep{2013ApJ...766...70S,2014ApJ...788L..35S,2019MNRAS.490.1743C}, HE 0435-1223 \cite{2019MNRAS.490.1743C,2017MNRAS.465.4895W},
SDSS 1206+4332 \cite{2019MNRAS.484.4726B}, WFI2033-4723 \citep{2020MNRAS.498.1440R}, PG 1115+080
\citep{2019MNRAS.490.1743C} and their sources cover the redshift range $0.654<z_s<1.789$. The time-delay distance and the luminosity distance for lensed quasar systems are summarized in Table. 2 of the Ref. \citet{2020MNRAS.498.1420W}.
For the earliest studied lens B1608+656, its $D_{\Delta t}$ measurement was represented by a skewed log-normal distribution. The $D_{\Delta t}$ of other lenses were given in the form of Monte Carlo Markov chains (MCMC) in a series of H0LiCOW papers. The distances posterior distributions are available on the H0LiCOW website \footnote{http://www.h0licow.org}.
The future Legacy Survey of Space and Time (LSST) of the Vera C. Rubin Observatory, would discover considerable amount of lensed quasar systems whose time delays could be measured \citep{2015ApJ...800...11L}. The project named ``Time Delay Challenge'' was proposed to test the measurability of SGL time delays with a blind analysis of mock LSST data \citep{2015ApJ...799..168D,2015ApJ...800...11L}. After testing many different algorithms, the results showed that time delay measurements of the precision and accuracy needed for time delay cosmography would indeed be possible \citep{2022arXiv220211101S,2020A&A...642A.193M,2020A&A...639A.101M,2018arXiv180101506D}.

\textit{Cosmographic approach to reconstruct the distance.---} In order to determine the curvature parameter and the Hubble constant, two dimensionless distances $d_l$ and $d_s$ are required. Other ingredients are directly measurable. In this work, we will reconstruct distances corresponding to redshifts  $z_l$ and $z_s$  by using the cosmographic approach to the data regarding the cosmic expansion history available from cosmic chronometers. Cosmographic techniques allow to study the evolution of the Universe in a model-independent way through kinematic variables. The procedure starts by introducing the series of cosmographic functions defined by subsequent time derivatives of the scale factor \citep{1972gcpa.book.....W,1998PThPh.100.1077C,2008cosm.book.....W,2003MNRAS.344.1057A,2004CQGra..21.2603V}.
First four of such functions are known as: the Hubble function, deceleration, jerk, snap:
\begin{equation}
H\equiv\frac{\dot{a}}{a},\,\, q\equiv\frac{-\ddot{a}}{aH^2},\,\,
j\equiv\frac{a^{(3)}}{aH^3},\,\, s\equiv\frac{a^{(4)}}{aH^4},\,\,
\end{equation}
where the dots represent cosmic time derivatives and $a^{(n)}$ stands
for the $n$-th time derivative of the scale factor.
With the above preparation, one can Taylor expand the Hubble parameter
\begin{eqnarray}\label{H-expansion} \nonumber
H(z)&=&H_0+\frac{dH}{dz}\big{|}_{z=0}z+\frac{1}{2!}\frac{d^2H}{dz^2}\big{|}_{z=0}z^2
+\frac{1}{3!}\frac{d^3H}{dz^3}\big{|}_{z=0}z^3+\cdots\\
&=&H_0\sum_{i=0}\mathcal{H}_z^iz^i,
\end{eqnarray}
and express the coefficients of the expansion in terms of cosmographic functions
\begin{eqnarray} \label{H-coefficients}
\mathcal{H}_z^0&=&1,\\ \nonumber
\mathcal{H}_z^1&=&1+q_0,\\ \nonumber
\mathcal{H}_z^2&=&\frac{1}{2}(j_0-q_0^2),\\ \nonumber
\mathcal{H}_z^3&=&\frac{1}{6}(-3j_0-4j_0q_0+3q_0^2+3q_0^3-s_0).\\ \nonumber
\end{eqnarray}
Here, the subscript "0" indicates the parameters at the present
epoch $(z = 0)$. In particular, $q_0<0$ indicates that the Universe is decelerating, and $q_0>0$ corresponds to an accelerating Universe. A positive sign of $j_0$ means a transition time between the deceleration and acceleration.

Furthermore, from (\ref{H-expansion}) and (\ref{H-coefficients}) one is able to build the dimensionless comoving distance $d(z)$  expansion,
\begin{equation} \label{D-expansion}
d_c(z)=\sum_{i=1}\mathcal{D}_z^iz^i,
\end{equation}
with
\begin{eqnarray} \label{D-coefficients}
\mathcal{D}_z^1&=&1,\\ \nonumber
\mathcal{D}_z^2&=&-\frac{1}{2}(1+q_0),\\ \nonumber
\mathcal{D}_z^3&=&\frac{1}{6}(2-j_0+4q_0+3q_0^2),\\ \nonumber
\mathcal{D}_z^4&=&-\frac{1}{24}(6-9j_0+18q_0-10j_0q_0+27q_0^2+15q_0^3-s_0). \nonumber
\end{eqnarray}
Using the above expressions, we can reconstruct the theoretical dimensionless comoving distance for the given strong lensing system at lens and source with the Taylor series expansion.

It has to be admitted that our work involves the data reaching high redshifts -- strong lensing systems up to the redshift of $z_s = 1.789$ and cosmic chronometers up to the redshift $z=2.3$.  Thus one might seriously worry about the convergence of the  cosmographic Taylor series.  One possible approach could be to consider auxiliary variables extending the convergence radii of standard Taylor expansions, such as $y=\arctan(z)$ and $y=1-1/(1+z)$ \citep{Aviles,Cattoen}, which leads to better convergence of these new expansions in the higher redshift range. However, such an approach is in a sense artificial -- new variables are losing the physical meaning. On the other hand, \citet{Gruberet} proposed the use of Pad\'{e} approximants (PA) for cosmographic analysis, which is much better than Taylor series expansion, as the convergence radius of PA is larger than Taylor Series expansion.

The $(n,m)$ order PA of a generic function $f(z)$ is its representation by a rational function
\begin{equation}
P_{n, m}(z)=\dfrac{\displaystyle{\sum_{i=0}^{n}a_{i} z^{i}}}{1+\displaystyle{\sum_{j=1}^{m}b_j z^{j}}}\,,
\label{eq:def_Pad}
\end{equation}
which agrees with $f(z)$ and all its derivatives up to $n+m$-th order at $z=0$.  As compared with the Taylor expansion, PA are known to have much better convergence, better approximate singular points and reduce uncertainty propagation. The comprehensive discussion on these issues can be found in \citep{2020MNRAS.494.2576C}. In particular it has been argued that the most stable order of the Pad\'{e} series is (2, 1). Therefore, we choose $P_{2,1}$ to approximate the Hubble parameter $H(z)$ in this work
\begin{equation}\label{coeff12}
H_{2,1}(z)=\frac{P_0+P_1z+P_2z^2}{1+Q_1z},
\end{equation}
where the $P_0$, $P_1$, $P_2$, and $Q_1$ are PA parameters. They also have no obvious physical meaning, but fortunately, one can always relate them to the physically relevant kinematic quantities $q_0$, $j_0$, $s_0$ and explicit expressions of this kind can be found in \citet{2019MNRAS.484.4484C}.
In the framework of PA, the dimensionless comoving distance $d(z)$ expansion at lensing redshift $z_l$ and source redshift $z_s$ also can be reconstructed by using this technique (see e.g. \citet{2020MNRAS.494.2576C}).

For the current data regarding the Hubble parameter, we turn to the newest compilation covering the redshift range of $0.01<z<2.3$
\citep{2003ApJ...593..622J,2012JCAP...07..053M,2015MNRAS.450L..16M,2016JCAP...05..014M},
which consists of 31 cosmic chronometer $H(z)$ data (denoted as CC) and 10 BAO $H(z)$ data. However, one should be aware that the BAO method relies slightly on cosmological models, i.e., the measurements based on the identification of BAO and the Alcock-Paczy\'{n}ski distortion from galaxy clustering depend on how ``standard rulers" evolve with redshift \citep{2012MNRAS.425..405B}. On the other hand, cosmic chronometers approach makes use of differential ages of passively evolving galaxies and thus is independent of any specific cosmological model. Hence the resulting $H(z)$ measurements are cosmological-model-independent \citep{LiuT20a,LiuT21a}. Therefore, in our work we consider only the cosmic chronometer data to reconstruct the $H(z)$ function and then derive dimensionless comoving distances ($d_l$ and $d_s$) within the redshift range of $0<z<2.3$, which covers the redshift of six  SGL systems very well.

Cosmic chronometers $H(z)$ data are sensitive to cosmographic parameters $\mathbf{p}=\{H_0, q_0, j_0 \}$ and corresponding  likelihood function $\mathcal{L}_{CC}$ is defined according to the following formula
\begin{eqnarray}
\mathcal{L}_{CC}=\prod_{i=1}^{31}\frac{1}{\sqrt{2\pi\sigma_{i}^2}}
\times\exp\bigg\{
-\frac{\left[H^{th}(z_i;\mathbf{p})
     - H^{obs}_{i}\right]^{2}}{2\sigma_{i}^{2}}\bigg\},
\end{eqnarray}
where the $H^{obs}_{i}$ is the $i-$th Hubble parameter measured by cosmic chronometers, with $\sigma_{i}$ representing its uncertainty, and the $H^{th}$ is theoretical Hubble parameter expressed in terms of cosmographic parameters, see Eqs. (9) and (14). It should be emphasized that convergence and goodness of fit decrease when higher-order terms are considered. We take into account in this paper is to analyze independently cosmographic expansion at third order with different technologies, i.e., Taylor series expansion denoted as $T_3$ and (2, 1) Pad\'{e} series expansion denoted as $P_{2,1}$.

In the case of SGL systems, time delay distance $D_{\Delta t}$ is the observable for which the likelihood was defined. For the lens B1608+656 the likelihood was given as a skewed log-normal distribution
\begin{equation}
\mathcal{L}_{D_{\Delta t}}=\frac{1}{\sqrt{2\pi}(x-\lambda_D)\sigma_D}\exp{
\Bigg[-\frac{((\ln(x-\lambda_D)-\mu_D)^2}{2\sigma_D^2}\Bigg]},
\end{equation}
with the parameters $\lambda_D=4000.0$, $\sigma_D=0.22824$, and $\mu_D=7.0531$,
where $x=D_{\Delta t}/$(1 Mpc).
For the other five lenses, the posterior distributions of $D_{\Delta t}$ were released in the form of Monte Carlo Markov Chains (MCMC) by the H0LiCOW team. A kernel density estimator was used to compute the likelihood $\mathcal{L}_{D_{\Delta t}}$ from the chains.
Finally, the log-likelihood of six lenses time-delay distances $\mathcal{L}_{D_{\Delta t}}$ and the log-likelihood of Hubble parameter, being independent, were used jointly to compute the posterior distributions of the cosmological parameters $(H_0, \Omega_k)$ and cosmography parameters $(q_0, j_0)$.
Hence, the log-likelihood sampled by using the Python MCMC module EMCEE \citep{2013PASP..125..306F} is given by
\begin{equation}
\mathrm{ln} {\cal L}= \mathrm{ln} ({\cal L}_{\rm{CC}})+\mathrm{ln}
({\cal L}_{D_{\Delta t}}).
\end{equation}
Concerning the priors on parameters, we employed a uniform prior on $H_0$ in the range [0, 150] $\Mpc$, $\Omega_k$ in the range [-1, 1], $q_0$ in the range [-2, 0.5], and $j_0$ in the range [-5, 5] to make sure that they are either physically motivated or sufficiently conservative.

\section{Results and discussion} \label{sec:method}

\begin{table*}
\renewcommand\arraystretch{1.5}
\caption{\label{tab:result} Summary of the constraints on the Hubble constant $H_0$, spatial curvature $\Omega_k$ and cosmographic parameters $q_0$ and $j_0$ in the framework of third order Taylor series and (2, 1) order Pad\'{e} expansions in cosmography by using time delay + cosmic chronometer observations.}

\begin{center}
\begin{tabular}{l| c c c c c c}
\hline
\hline
Parametric order  & $H_0$ ($\Mpc$) &$\Omega_k$ & $q_0$ & $j_0$ & $\chi^2_{dof}$ & BIC  
\\
\hline
Taylor $T_3$ & $72.24^{+2.73}_{-2.52}$ & $0.062^{+0.117}_{-0.078}$ & $ -0.645^{+0.126}_{-0.124}$& $0.901^{+0.241}_{-0.225}$   & $1.43$ & $65.98$  \\
\hline

Taylor $T_3$ (fixed $\Omega_k$ )  & $70.47^{+1.14}_{-1.15}$ & $-$ & $ -0.560^{+0.063}_{-0.061}$& $0.746^{+0.122}_{-0.114}$  & $1.43$ & $62.30$ \\
\hline
Taylor $T_3$ (fixed $H_0$ )  & $-$ & $0.123^{+0.060}_{-0.046}$ & $ -0.716^{+0.051}_{-0.051}$& $1.024^{+0.127}_{-0.120}$ & $1.44$ &$62.75$ \\
\hline
Pad\'{e} $P_{2,1}$  & $  72.45^{+1.95}_{-2.02}$ & $ 0.069^{+0.116}_{-0.103}$ & $ -0.822^{+0.115}_{-0.218}$& $1.161^{+0.237}_{-0.204}$ &$1.39$ & $64.19$ \\
\hline
Pad\'{e} $P_{2,1}$ (fixed $\Omega_k$ )  & $71.66^{+1.15}_{-1.57}$ & $-$ & $ -0.795^{+0.100}_{-0.081}$& $ 1.128^{+0.122}_{-0.279}$ &$1.42$ &$62.18$\\
\hline
Pad\'{e} $P_{2,1}$ (fixed $H_0$ )  & $-$ & $0.101^{+0.090}_{-0.072}$ & $  -0.847^{+0.056}_{-0.072}$& $1.179^{+0.109}_{-0.110}$ &$1.38$ &$60.58$  \\
\hline
\hline
\end{tabular}
\end{center}
\end{table*}

\begin{figure}
\centering
\includegraphics[scale=0.55]{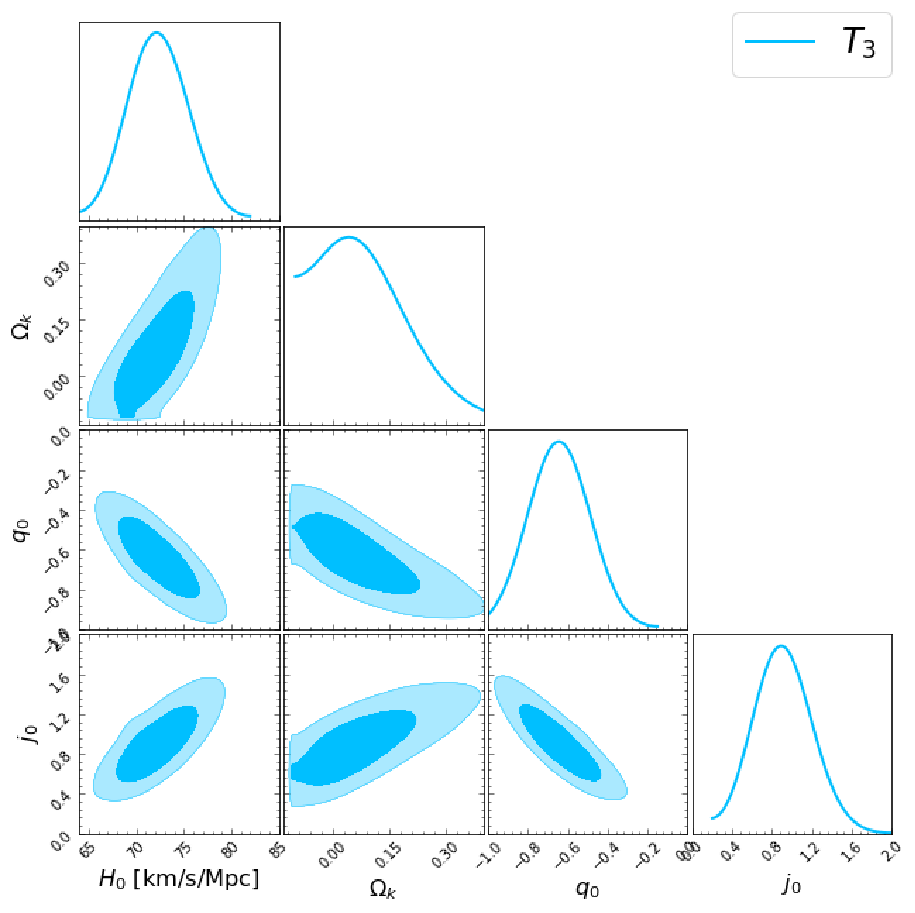}
\caption{ The 1D marginalized probability distributions and 2D regions with $1\sigma$ and $2\sigma$ contours corresponding to these parameters $(H_0, \Omega_k, q_0, j_0)$ in the third order Taylor series expansion cosmography, constrained by the six time delay measurements and cosmic chronometer data.}
\label{fig1}
\end{figure}

\begin{figure*}
\begin{center}
\includegraphics[scale=0.5]{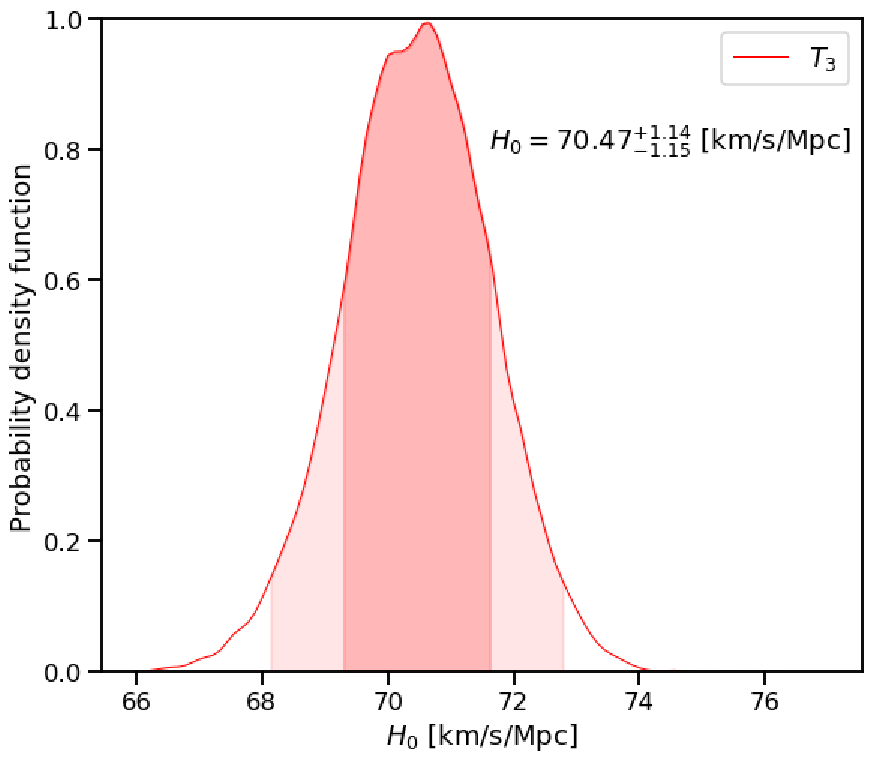}
\includegraphics[scale=0.5]{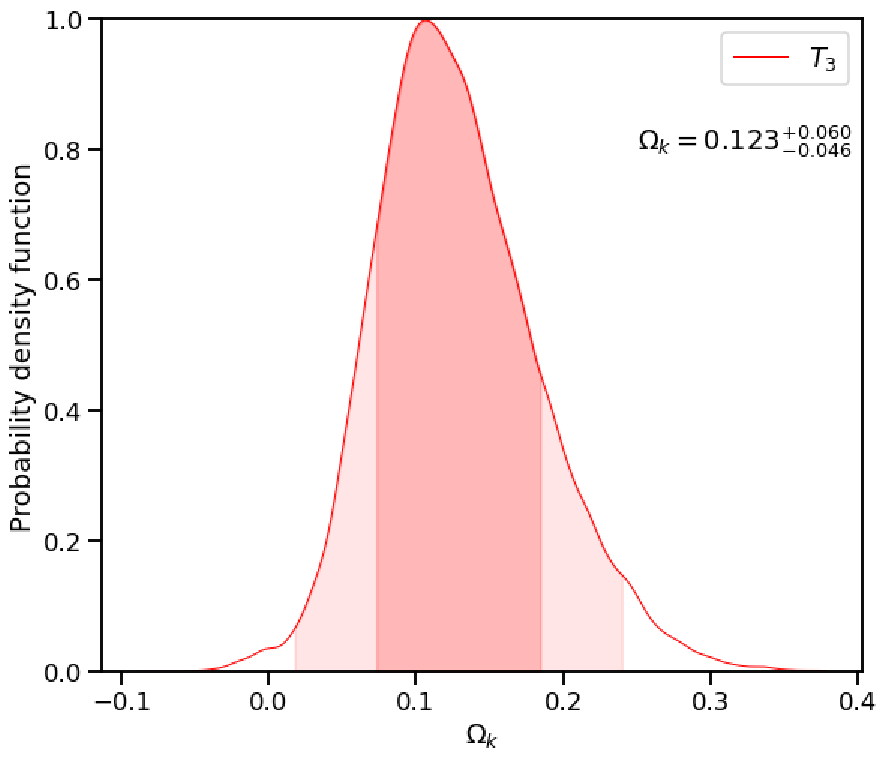}
\end{center}
\caption{\textit{Left:} The posterior probability density function of the Hubble constant in the third order Taylor series expansion in a flat Universe.
\textit{Right:} The posterior probability density function of the spatial curvature parameter in the third order Taylor series expansion with fixed $H_0=73.24\Mpc$.  }
\label{fig2}
\end{figure*}

\begin{figure}
\centering
\includegraphics[scale=0.65]{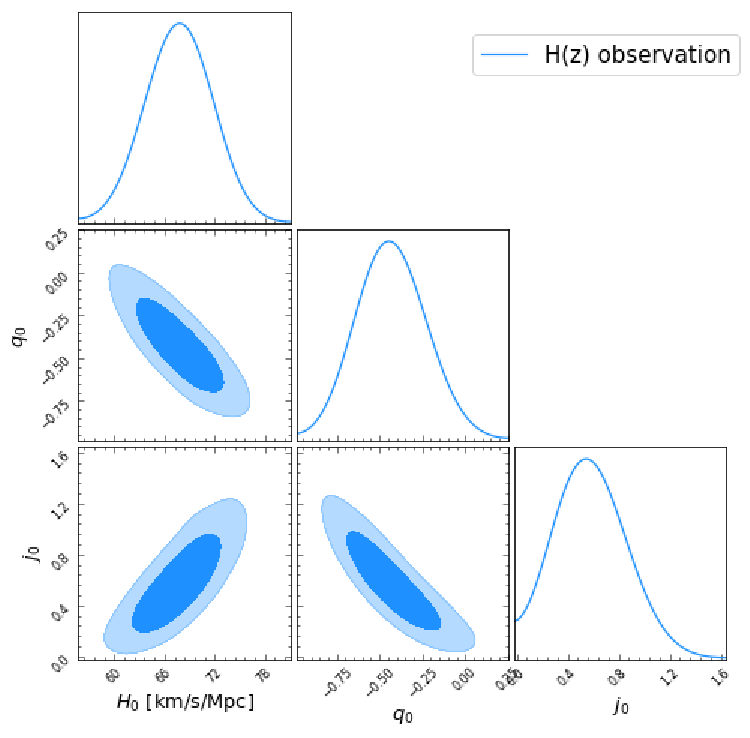}
\caption{The 1D marginalized probability distributions and 2D regions with 1$\sigma$ and 2$\sigma$ confidence level for parameters ($H_0, q_0, j_0$) in the framework of the third order Taylor series expansion cosmography by using alone cosmic chronometer data. }
\end{figure}

Firstly, we consider the Hubble parameter which using third order of Taylor expansion in cosmography. The 1D marginalized probability distributions and 2D regions with $1\sigma$ and $2\sigma$ contours corresponding to these parameters $(H_0, \Omega_k, q_0, j_0)$, constrained by the six time delay measurements and cosmic chronometer data, are displayed in Figure 1. The median values plus the distance to $16^{th}$ and $84^{th}$ percentiles for the four parameters
are $H_0=72.24^{+2.73}_{-2.52}\Mpc$, $\Omega_k=0.062^{+0.117}_{-0.078}$,
$q_0=-0.645^{+0.126}_{-0.124}$, and $j_0=0.901^{+0.241}_{-0.225}$. Our results support the flat Universe at $1\sigma$ confidence level, and the central value of $H_0$ is in agreement with the result inferred from H0LiCOW collaboration, which demonstrates the consistency and validity of our method. This clear result from available lensed quasars is also consistent with recent analysis of distance ratios in strong lensing systems \citep{2017ApJ...834...75X,2019PhRvD.100b3530Q,2020ApJ...901..129L,2020ApJ...889..186Z,LiuT20b,LiuT21b}, which is a technique complementary to SGL time-delay.
\begin{figure}
\centering
\includegraphics[scale=0.55]{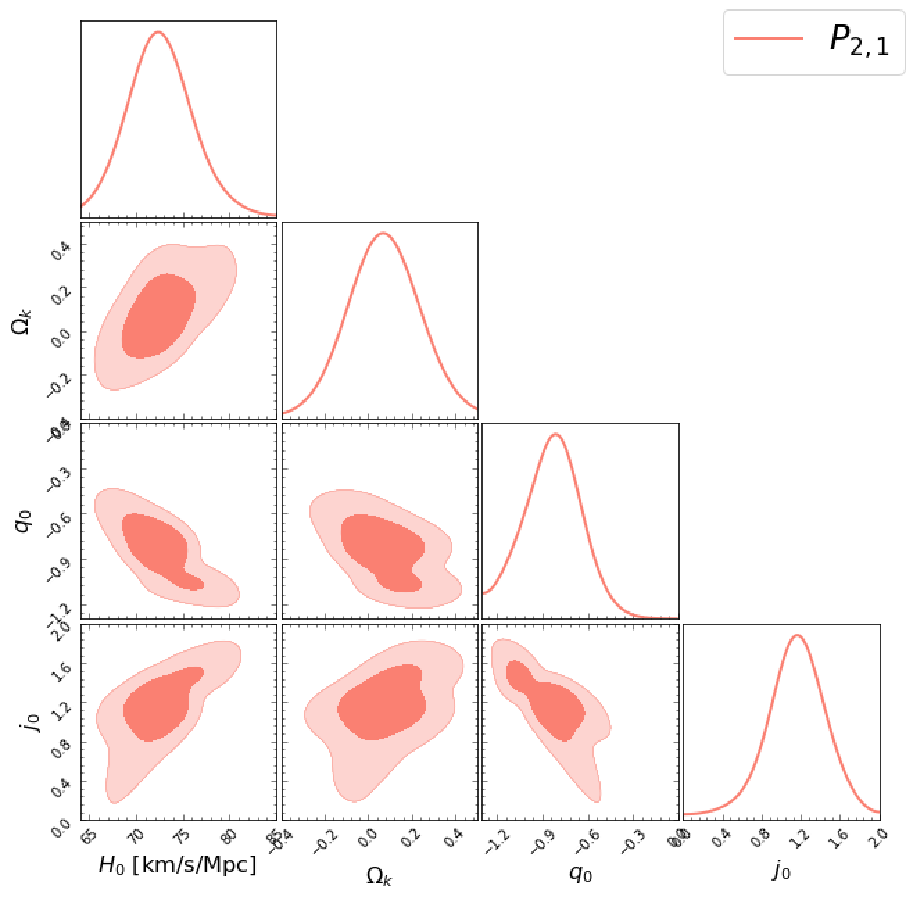}
\caption{ The 1D marginalized probability distributions and 2D regions with $1\sigma$ and $2\sigma$ contours corresponding to these parameters $(H_0, \Omega_k, q_0, j_0)$ in the (2, 1) order Pad\'{e} expansion, constrained by the six time delay measurements and cosmic chronometer data}.
\label{fig1}
\end{figure}

\begin{figure*}
\begin{center}
\includegraphics[scale=0.5]{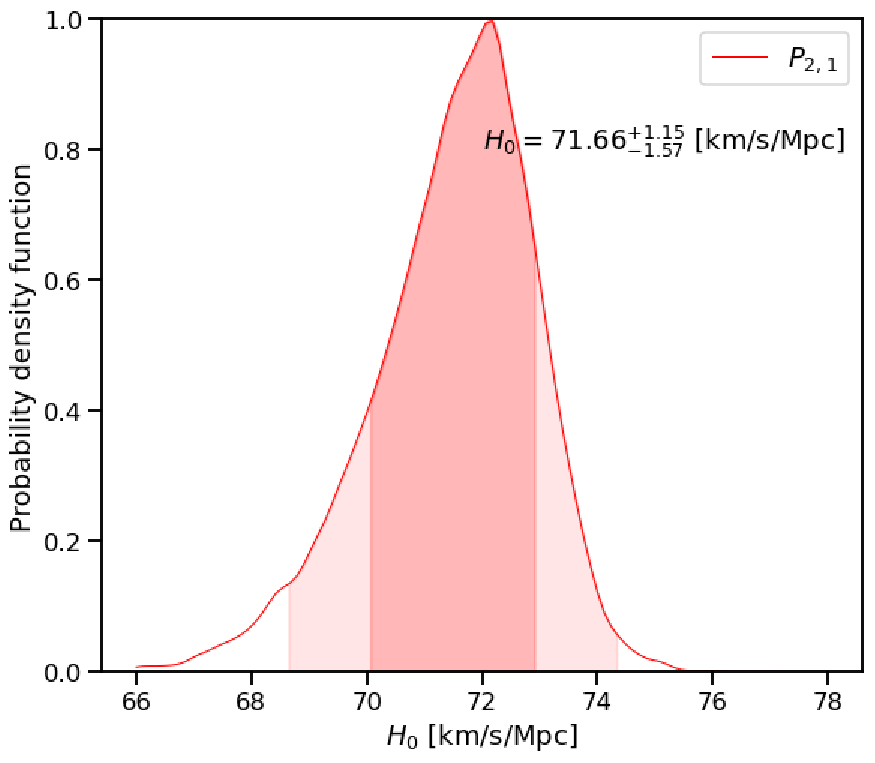}
\includegraphics[scale=0.5]{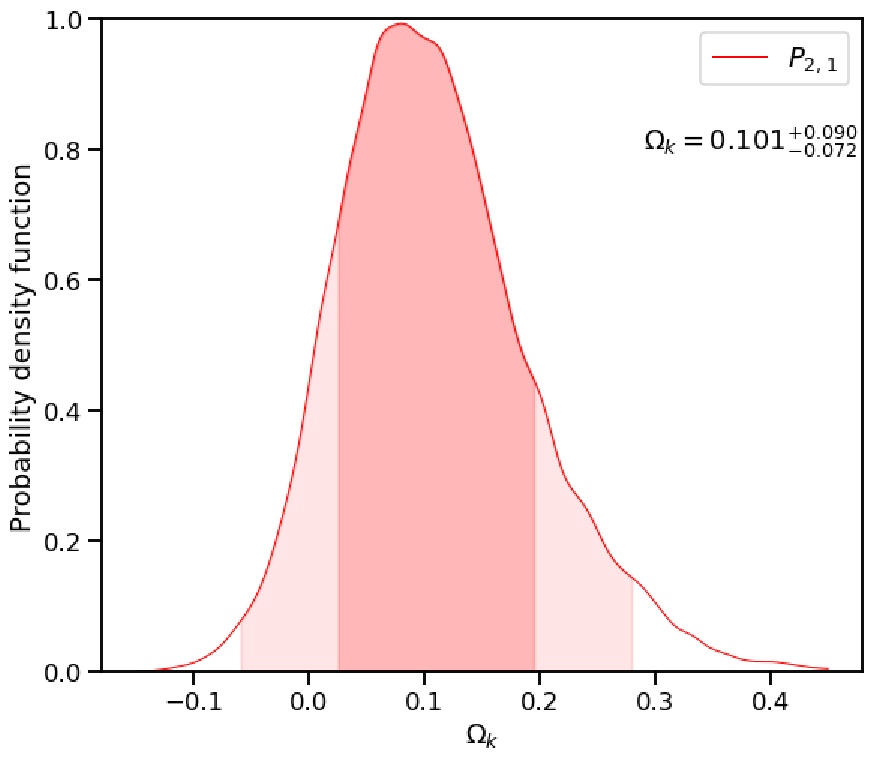}
\end{center}
\caption{The same as Figure 2, but using the (2, 1) order Pad\'{e} expansion. }
\label{fig2}
\end{figure*}

If we assume a flat Universe, one can clearly see from Table 1 and the left panel of the Figure 2 that $H_0=70.47^{+1.14}_{-1.15}$ $\Mpc$, which falls between the SH0ES and \textit{Planck} CMB results. The detailed numerical results from joint
time delays + cosmic chronometers analysis are displayed in rows 1 and 2 of Table 1 for a non-flat and flat Universe, respectively.
In particular, the cosmographic coefficients do not differ much between these two cases. The deceleration parameters $q_0=-0.645^{+0.126}_{-0.124}$ and $q_0=-0.560^{+0.063}_{-0.061}$  corresponding to the non-flat and flat cases, respectively, are mutually consistent. Their values indicate an accelerating expansion of the Universe. To better understand the meaning of these  parameters, let us consider a flat $\Lambda$CDM model. Then, one can relate cosmographic parameters to the physical quantity $\Omega_{m0}$ (i.e. the matter density parameter at present time) according to $q_0=-1+3/2\Omega_{m0}$ and  $j_0=1$. It is easy to check that our results are compatible with a flat $\Lambda$CDM model with the current matter density parameter $\Omega_{m0}=0.311$ obtained from \textit{Planck} CMB observations within 1$\sigma$ confidence level. In the work of \citep{2020MNRAS.494.2576C}, the authors have discussed the role of spatial curvature in the cosmographic context. Degeneracy among coefficients $(q_0, j_0)$ and spatial curvature appears due to the fact that all cosmographic parameters are related to the Hubble function $H(z)$. In order to highlight the influence of spatial curvature on cosmographic parameters, in another scenario we fixed the value of the Hubble constant as $H_0=73.24\Mpc$. We adopted the Gaussian kernel density estimator to compute probability density function of the spatial curvature parameter $\Omega_k$ in the right panel of Figure 2. Full numerical results regarding the cosmographic parameters are given in the row 3 of Table 1. In this case, our findings support an open Universe at $2.6 \sigma$ confidence level. Moreover, our results also confirm the conclusions of \citep{2020MNRAS.494.2576C}.

In order to clearly demonstrate and compare the contribution from each probe on parameter constraints in our work, we note that using gravitational lensing alone for time-delay observations requires assuming a particular cosmological model, we thus only using cosmic chronometer data to constrain $H_0,$ $q_0$ and $j_0$. In the framework of the third order Taylor series expansion cosmography, Figure 3 shows the 1D marginalized probability distributions and 2D regions with 1$\sigma$ and 2$\sigma$ confidence level. As one can see, the median values plus the distance to $16^{th}$ and $84^{th}$ percentiles for the three parameters
are $H_0=67.782^{+3.09}_{-3.25}\Mpc$, $q_0=-0.437^{+0.176}_{-0.158}$, and $j_0=0.564^{+0.256}_{-0.224}$. Compared with the results showed in Figure 1, this shows that our work can not only improve the precision of $H_0$ constraint, but also improve the precision of kinematic parameters. More importantly, our cosmological model-independent approach can simultaneously determine the Hubble constant, cosmic curvature and kinematic parameters (using $H(z)$ alone does not constrain on cosmic curvature). In other word, the time-delay of the gravitational lens improves the precision of the constraint on $H_0$, and the cosmic chronometer data improves the precision of the constraint of the cosmic kinematic parameters.

Results for the (2, 1) order Pad$\'{e}$ approximant used for cosmographic analysis are reported in row 5 of Table 1 and displayed graphically in Figure 4.
Comparing the obtained value of $H_0=72.45^{+1.95}_{-2.02}\Mpc$ with  $H_0=72.24^{+2.73}_{-2.52}$ inferred from Taylor third order expansion, we see very little effect on the results coming from different reconstruction methods. Most likely this is so, because the inferred value of $H_0$ is dominated by the contribution of very precise data regarding gravitational lensing time delays. Regarding the cosmic curvature parameter, our results show that there is almost no difference between the curvature parameters obtained by the two reconstruction methods, which once again proves the effectiveness of our methods and techniques. However, the PA method yields the smaller value of deceleration and the higher value of jerk parameter: $q_0= -0.822^{+0.115}_{-0.218}$ and $j_0=1.161^{+0.237}_{-0.204}$, respectively. This result is likely due to strong degeneracy between the cosmographic parameters themselves. Therefore, we expect to seek for other types of astronomical data samples that might break the degeneracy.

Assuming zero spatial curvature, we obtain $H_0=71.66^{+1.15}_{-1.57}$ $\Mpc$ (see the posterior probability density function in left panel of the Figure 5; representing
a precision of 2.2\%), $q_0=-0.795^{+0.100}_{-0.081}$, and $j_0=1.128^{+0.122}_{-0.279}$, which also fall between the SH0ES and Planck CMB results.
Similarly to the third order Taylor expansion, we fix the value of the Hubble constant as $H_0=73.24$ $\Mpc$ to explore the spatial curvature, the curvature parameter of posterior probability density function under this assumption is shown in the right panel of  Figure 5. An open Universe is also supported at
$1.3\sigma$ confidence level. The differences between  PA derived cosmological parameters and those obtained with Taylor expansion method are consistent within $1\sigma$ uncertainties.

In order to quantify the relative performance of Taylor expansion and PA, we use the Bayes Information Criterion (BIC), which is a well known tool of model selection theory \citep{BurnhamAnderson}. It is defined as
\begin{equation}
BIC=-2\ln {\cal L}_{max}+k\ln N,
\end{equation}
where $\mathcal{L}_{max}$ is the maximum value of the likelihood, $N$ denotes the number of data points, and $k$ represents the number of free parameters. The BIC values for Taylor and Pad\'{e} expansions are listed in Table 1, together with reduced chi-square  $\chi^2_{dof}$, i.e. $\chi^2_{dof}=\chi^2_{max}/36$ with degrees of freedom (31+6-1). According to these results, we
conclude that Pad\'{e} expansion is slightly preferred over the Taylor expansion with the BIC difference $\Delta BIC=1.79$. An evidence for a mild difference between the support given to competing models by the data starts with the difference $\Delta BIC\geq 2$ and can be regarded strong with $\Delta BIC\geq 6$ \citep{1978AnSta...6..461S}. Hence, we conclude that adopting different extension techniques has only a minimal impact on our results.

On the other hand, while the best fitting values of the cosmographic parameters $q_0, j_0$ and Hubble constant $H_0$ are known, we can reconstruct the cosmic expansion history. In Figure 6, we show the cosmic expansion history based on the third order fitting in cosmography, which is consistent well with cosmic chronometer observations. This is already a significant result. In the future,
when the number and quality of the data improve, one can expect that our approach will yield much more precise and accurate
reconstruction of cosmic expansion history over a wide range of redshifts as well as determination of the Hubble constant and cosmic curvature.

\section{Conclusion} \label{sec:res}

\begin{figure}
\centering
\includegraphics[scale=0.5]{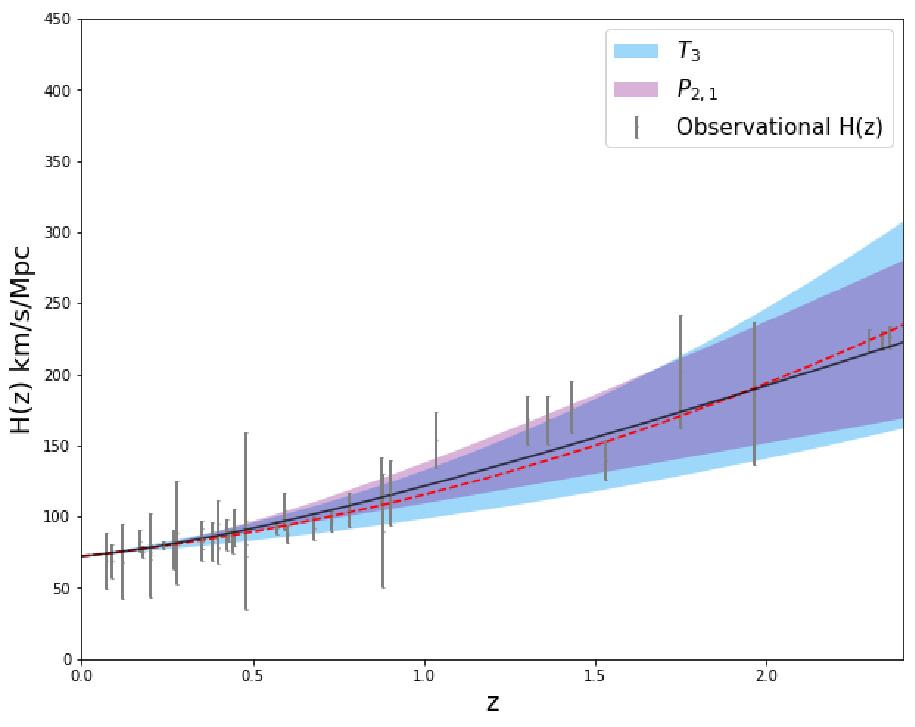}
\caption{Reconstructed Hubble expansion history by using the third order Taylor series and (2, 1) order Pad\'{e} expansions.  }
\label{fig4}
\end{figure}

The Hubble constant and spatial curvature of the Universe are still the most important cosmological parameters in modern cosmology.
In this work, we used a model-independent approach to determine $H_0$ and $\Omega_k$
simultaneously by analysing cosmic chronometers data in combination with time delay measurements of six well studied SGL systems.
Unlike many recent applications of various cosmological probes based on the polynomial expansion, our method used cosmography to reconstruct distances from cosmic chronometer observations. Beyond a very general basic assumption of homogeneity and isotropy of the Universe (independence of the details of cosmological model), the advantage of cosmography is that parameters like $H_0$, $q_0$, $j_0$ and $s_0$ can be directly obtained from the data without need of calibrating the intercept (nuisance parameter) like in the case of standard candles or standard rulers.

Firstly, in the framework of the third order Taylor expansion, our results showed that the mean value of $H_0$ is in agreement with the value inferred by H0LiCOW collaboration, and a flat Universe is supported within $1 \sigma$ confidence level based on current observational data. This is consistent with the previous results given in \citet{2019PhRvL.123w1101C,2020ApJ...897..127W,2021MNRAS.503.2179Q}, in which the authors respectively used the Pantheon SNIa, the non-linear relation between the ultraviolet and X-ray fluxes in quasars, and angular size vs. redshift relation from compact structures in radio quasars combined with time delay measurements of SGL to obtain the $H_0$ and $\Omega_k$. However, their methods usually required the reconstruction of cosmological distances, phenomenologically modeled by some  polynomials.  This approach often involves problems about the physical meaning of polynomial coefficients and, what is more important, the calibration of ``nuisance parameter". Calibration introduces additional systematic uncertainties, hence one may claim that our results regarding $H_0$ and $\Omega_k$ are more accurate than these previous assessments. The deceleration parameters turned out to be  $q_0=-0.645^{+0.126}_{-0.124}$ which indicates an accelerating the expansion of the Universe. Our results are compatible, within 1$\sigma$ confidence level, with a flat $\Lambda$CDM model having the current matter density parameter $\Omega_{m0}=0.311$ as suggested from \textit{Planck} CMB observation.

Assuming a flat Universe, we obtained $H_0=70.47^{+1.14}_{-1.15}$ $\Mpc$, which falls between the SH0ES and \textit{Planck} CMB data results.
In order to highlight the influence of spatial curvature on cosmographic parameters, we fixed the value of Hubble constant as $H_0=73.24\Mpc$. Under this assumption, our findings supported an open Universe at $2.6 \sigma$ confidence level. This means that there is a positive correlation between the Hubble constant and the spatial curvature parameter. Moreover, our results also confirm the conclusions formulated in \citep{2020MNRAS.494.2576C}. It should be stressed, however, that our work offered a new approach to constrain both the spatial curvature and cosmographic parameters. Although the degeneracy between them still exists, our method alleviates it to some extent in comparison with the cosmography alone.
Considering that redshift coverage of the data we used extends to $z \sim 2.3$, the issue of convergence regarding the Taylor series cosmographic expansion becomes important. Therefore we also used the technique of Pad\'{e} approximants, in particular the (2,1) PA known to perform the best in cosmography \citep{2020MNRAS.494.2576C}. As discussed above, the results regarding $H_0$ and cosmic curvature were similar as in Taylor expansion, while deceleration and jerk parameters turned to be noticeably different. The BIC criterion used to quantify the performance of two competing reconstruction techniques yielded a weak preference of PA over Taylor expansion. This result is however inconclusive.

As a final remark, there are many potential ways to improve our method. For instance, current and future surveys like the Dark Energy Survey (DES) \citep{2018MNRAS.481.1041T}, the Hyper SuprimeCam Survey \citep{2017MNRAS.465.2411M}, and the 
Legacy Survey of Space and Time (LSST)
\citep{2010MNRAS.405.2579O,2015ApJ...811...20C} will bring us hundreds of thousands of lensed quasars in the most optimistic discovery scenario. Even if only some small fraction of them will have precise measurements of time delays between multiple images, the resulting statistics will outshine current catalogs. With high-quality auxiliary observations, one can use high-cadence, high-resolution and multi-filter imaging of the resolved lensed images, to derive accurate determination of the Fermat potential, which will increase the precision of time delay distance by an order of magnitude. At the same time, we also expect that following the surveys like e.g.
WiggleZ Dark Energy Survey \citep{2010MNRAS.401.1429D}, future observations of cosmic chronometers will be greatly improved. It is reasonable to expect that our approach will play an increasingly important role in the simultaneously constraining both Hubble constant and the spatial curvature, with future surveys of SGL systems and observations of cosmic chronometer.

\acknowledgments

This work was supported by the National Natural Science Foundation of China under Grants Nos. 12203009, 12021003, 11690023, and 11920101003; the National Key R\&D Program of China No. 2017YFA0402600; the Strategic Priority Research Program of the Chinese Academy of Sciences, Grant No. XDB23000000; the Interdiscipline Research Funds of Beijing Normal University; the China Manned Space Project (Nos. CMS-CSST-2021-B01 and CMS-CSST-2021-A01); and the CAS Project for Young Scientists in Basic Research under Grant No. YSBR-006. M.B. was supported by Foreign Talent Introducing Project and Special Fund Support of Foreign Knowledge Introducing Project in China (No. G2021111001L).


\end{document}